# EURONEAR – FIRST LIGHT CURVES AND PHYSICAL PROPERTIES OF NEAR EARTH ASTEROIDS

A. AZNAR MACÍAS[1], M. PREDATU[2], O. VADUVESCU[3], J. OEY[4]

[1]Isaac Aznar Observatory, Centro Astronómico del Alto Turia, Aras de los Olmos, Valencia, Spain,
E-mail: aptog@aptog.com
[2]University of Craiova, Department of Physics, A. I. Cuza 13, Craiova, RO-200585, Romania,
[3]Isaac Newton Group, Apt. de correos 321, E-38700, Santa Cruz de La Palma, Canary Islands, Spain,
E-mail: ovidiu.vaduvescu@gmail.com
[4] Blue Mountain Observatory, 94 Rawson Pde. Leura, NSW 2780, Australia,
E-mail: Julianoey1@optusnet.com.au
*Correspondent author*: predatumarian@yahoo.com

*Abstract*. Part of the *European Near Earth Asteroids Research* (EURONEAR) project, in 2014 we started a survey to observe light curves of *Near Earth Asteroids* (NEAs) using the available telescopes of this network. This is the first paper presenting light-curves and physical properties of 17 NEAs observed by two amateur astronomers owning small facilities located in good sites, namely a 0.36 m telescope at Isaac Aznar Observatory (Aras de los Olmos, Spain) and a 0.61 m telescope at Blue Mountain Observatory (Leura, Australia). We confirm most recent or older results and find new ones.

## 1. INTRODUCTION

*Near Earth Asteroids* (NEAs) represent a particular important class of minor planets due to their potential hazard threat to Earth, therefore their physical studies are extremely important. NEAs are defined based on their orbits, as objects approaching the Sun (perihelion distance $q$) less than 1.3 astronomical units (au).

Based on their orbital distance to the Sun (semi-major axis $a$, perihelion distance $q$ and aphelion distance $Q$), NEAs are divided in three main subgroups: Apollo ($a \geq 1.0$ au, $q \leq 1.017$ au), Amor ($a \geq 1.0$ au, $1.017 < q < 1.3$ au) and Aten ($a < 1.0$ au, $Q \geq 0.983$ au) [1].

Currently we know more than 14,000 NEAs [1], 99% discovered by six U.S.-lead surveys (in historic order: Spacewatch, LONEOS, NEAT, LINEAR, Catalina and Pan-STARRS). Nevertheless, only extremely few NEAs (about 7% or one thousand objects) have been characterized photometrically *via* light curves, due to the lack of dedicated professional telescopes, and we remind here the efforts of P. Pravec *et al*. [2, 3, 4, 5] and A. Galad *et al.* [6] using 1 m class and smaller telescopes (sometime during coordinated campaigns), and those of T. Kwiatkovski *et al*. who used the giant SALT 10 m telescope to characterize only very small objects and fast rotators [7, 8]. In the meantime, dedicated amateur astronomers could bring serious contribution using their smaller telescopes (mostly 0.3–0.6 m) for longer time, preferably located in decent sites. Especially, this setup it is used to characterize larger objects which need few continuing nights to resolve. We remind the efforts of the North-American amateurs Brian Warner and Robert D. Stephens who brought significant contributions since 1999, observing and publishing photometry of about 500 NEAs from the actual number of about 1,000 NEA saving valid curves (uncertainty factor U $\geq$ 2-, acc. to B. Warner).

Since 2006, the *European Near Earth Asteroids Research* (EURONEAR) project is aiming to contribute to the knowledge of orbital and physical characterization of NEAs part of an European network which includes 14 nodes from 7 European countries plus Chile [9], comprising since 2015 three nodes from Romania. In this context, since 2014, EURONEAR has observed about 100 NEA light curves (most having no photometric data) using 15 telescopes available to the network, with diameters between 0.35 m and 4.2 m.

This article presents the first brighter NEAs observed with two small private amateur telescopes, while other subsequent four papers will present fainter NEAs observed with larger telescopes. Two main objectives has been considered: to determine the equivalent diameters and the preliminary shape of the observed NEAs, based on the light curve analysis. Most of the targets presented in this paper have been previously or very recently observed by few other authors, raising the opportunity to compare our data with independent literature results, validating our reduction techniques to be applied in future papers.

## 2. OBSERVATIONS

### 2.1. THE OBSERVATORIES AND TELESCOPES

#### 2.1.1. The *Isaac Aznar Observatory*

This private observatory is owned by the Spanish amateur astronomer Amadeo Aznar who has contributed the great majority amount of data for this paper. It is located in the *Centro Astronómico del Alto Turia*, *Aras de los Olmos*, Valencia, Spain, at 1270 meters above the sea level, in one of the darkest night-sky of Iberian Peninsula (limiting magnitude 22.1 mag/arcsec$^2$). The optical system consists in a 0.36 m diameter Schmidt – Cassegrain F/10 Meade LX200 telescope which can be remotely controlled. The CCD camera is a SBIG STL 1001e with adaptive optics and CCD sensor 1024 × 1024 pixels of size 24.6 μm, resulting in 1.44 "/pixel and square 25' field. The camera holds broad band Astrodon *Sloan r* and *J. V.* filters, working at –20 C° with cooled water in the summer. Observatory website: http://www.apt.com.es/web/index.php/observatorio.html.

### 2.1.2. The *Blue Mountain Observatory*

This private observatory is owned by the Australian amateur astronomer Julian Oey, being located at close driving distance of Sydney at 946 meters altitude, under a dark sky 21.4 mag/arcsec$^2$. The optical system consists in a 0.61 m CDK Plane Wave F/6.5 telescope which could be remotely controlled. The camera consists in a 2048 × 2048 CCD with 13.5 μm pixels resulting in a scale of 0.70"/pixel and square 24' field, being endowed with a broad band Johnson *R* filter. The observatory website is http://www.bluemountainsobservatory.com.au.

### 2.2. THE TARGETS

In Table 1 we list the 17 observed NEAs ordered by the asteroid number in the left column, followed by the orbital class, published apparent magnitude *H*, albedo $\rho$ [10, 11, 12], STC spectral type class [13], observing night(s) interval, proper motion μ (in "/min), exposure time (in sec) and number of exposures taken.

In Table 2 we include the derived physical data of the 17 observed NEAs. We list the asteroid number, the fitted rotation period *P*, reduced magnitude *HR*, the light curve amplitude *A*(0) at zero phase angle (in mag), the modeled ellipsoid axis *a*, *b* and *c* (all in km) and the ratios *a/b* and b/c.

We did not constrain any parameter while planning the observations the main condition being the lack of photometric data (not secured light curve at the time). Based on orbital class, 8 targets are Amors, 6 Apollos and 3 are Athens. Based on absolute magnitudes, most observed targets (9 objects) have *H* in the 17$^{th}$ range, while the rest are brighter (6 objects with 11.0 < *H* < 16.5) or fainter (only 2 objects with 17.5 < *H* < 19.0). Most targets (12 objects) have physical sizes in the 1–3 km range. Five targets of all are larger, all being accessible to small telescopes.

*Table 1*

Literature and observational data of the 17 Near Earth Asteroids (NEAs) observed from the Isaac Aznar Observatory (most objects) and Blue Mountain Observatory (only object 141354)

| Ast. Nr. | Class | H | Albedo ρ | STC | Obs. nights | μ | Exp. time | Nr. Exp |
|---|---|---|---|---|---|---|---|---|
| 433 | Amor | 11.2 | 0.25 | S | 18/07–17/07/16 | 0.50 | 30 | 434 |
| 1866 | Apollo | 12.4 | 0.37 | S | 26/04–03/05/16 | 1.03 | 180 | 172 |
| 1980 | Amor | 13.9 | 0.13 | S | 15/10–23/10/15 | 1.68 | 180 | 113 |
| 4055 | Amor | 14.7 | 0.33 | V | 28/08–09/08/15 | 4.06 | 180 | 298 |
| 7350 | Apollo | 17.2 | 0.05 | – | 24/04–25/04/16 | 2.72 | 300 | 78 |
| 10150 | Amor | 15.3 | 0.15 | – | 23/06–26/06/16 | 2.96 | 300 | 144 |
| 35396 | Apollo | 16.9 | 0.1–0.2 | S | 22/04–24/04/16 | 2.26 | 300 | 102 |
| 68216 | Apollo | 16.5 | 0.29 | – | 06/03–10/03/16 | 1.15 | 200 | 68 |
| 68346 | Apollo | 16.8 | 0.1–0.2 | S | 17/05–27/05/16 | 1.63 | 300 | 98 |
| 85989 | Aten | 17.1 | 0.07 | K | 06/07–08/07/16 | 1.82 | 120 | 215 |
| 137805 | Aten | 16.6 | 0.02 | S | 31/01–02/02/16 | 1.64 | 160 | 117 |
| 141354 | Amor | 17.3 | 0.15 | – | 03/05–04/05/16 | 1.99 | 180 | 84 |
| 154244 | Amor | 17.5 | 0.1–0.2 | S | 22/06–30/06/16 | 2.09 | 180 | 156 |
| 163899 | Aten | 17.3 | 0.1–0.2 | S | 29/11–04/12/15 | 2.31 | 240 | 67 |
| 331471 | Apollo | 15.6 | 0.1–0.2 | – | 31/05–30/06/16 | 10.74 | 90 | 1140 |
| 337866 | Amor | 18.7 | 0.1–0.2 | S | 08/02–21/02/16 | 2.10 | 240 | 110 |
| 385186 | Amor | 17.4 | 0.1–0.2 | L | 22/07–27/07/15 | 6.22 | 180 | 253 |

*Table 2*

Photometric and Derived physical data of the 17 observed Near Earth Asteroids (NEAs). Asteroid number, period, reduced magnitude*, Amplitude in zero phase angle, a axis (km), b axis (km), c axis (km), a/b semi-axis ratio, b/c semi-axis ratio

| Ast. Nr. | P | HR* | A(0) | a | b | c | a/b | b/c |
|---|---|---|---|---|---|---|---|---|
| 433 | 5.27 | 11.7 | 0.252 | 23.3 | 18.47 | 18.22 | 1.261 | 1.014 |
| 1866 | 2.40 | 16.0 | 0.056 | 7.21 | 6.85 | 6.78 | 1.053 | 1.009 |
| 1980 | 7.23 | 15.1 | 0.278 | 5.99 | 4.64 | 4.35 | 1.292 | 1.065 |
| 4055 | 7.47 | 14.7 | 0.395 | 2.16 | 1.50 | 1.37 | 1.438 | 1.091 |
| 7350 | 3.93 | 17.0 | 0.089 | 2.15 | 1.99 | 1.95 | 1.085 | 1.018 |
| 10150 | 2.96 | 16.6 | 0.131 | 4.25 | 3.77 | 3.66 | 1.128 | 1.028 |
| 35396 | 3.15 | 15.5 | 0.431 | 1.30 | 0.88 | 1.12 | 1.487 | 0.782 |
| 68216 | 2.49 | 16.0 | 0.160 | 2.49 | 2.15 | 2.09 | 1.159 | 1.027 |
| 68346 | 5.63 | 17.5 | 0.472 | 1.56 | 1.01 | 1.23 | 1544 | 0.818 |
| 85989 | 7.65 | 16.0 | 0.609 | 1.84 | 1.05 | 0.50 | 1.753 | 2,075 |
| 137805 | 3.47 | 15.5 | 0.043 | 3.87 | 3.72 | 3.69 | 1.040 | 1.008 |
| 141354 | 19.84 | 15.7 | 0.338 | 1.66 | 1.22 | 1.08 | 1.365 | 1.126 |
| 154244 | 4.61 | 14.8 | 0.396 | 1.13 | 0.79 | 0.86 | 1.441 | 0.912 |
| 163899 | 173.40 | 15.5 | 0.910 | 1.03 | 0.45 | 0.76 | 2.311 | 0.579 |
| 331471 | 45.90 | 16.0 | 0.189 | 3.12 | 2.63 | 2.47 | 1.190 | 1.062 |
| 337866 | 6.33 | 11.7 | 0.063 | 0.85 | 0.80 | 0.79 | 1.060 | 1.012 |
| 385186 | 2.49 | 16.0 | 0.065 | 0.55 | 0.52 | 0.51 | 1.062 | 1.014 |

* Reduced magnitude corrected to unity distances at observed phase angle and assuming that the asteroid is at 1 AU from Sun and 1 AU from Earth [14].

## 3. DATA REDUCTION

### 3.1. IMAGE PROCESSING

Typical bias, flat and dark frames were used for image processing using the CCD Soft provided by Software Bisque. Operations like setting multiple groups of various combinations of bias, dark and flat-field frames were performed on the raw image. Also operation like averaging or median combine bias were performed on dark frame subtract and flat-field images.

### 3.2. LIGHT CURVE REDUCTION

The photometric measurements and period analysis were conducted using the *MPO Canopus* 10.7.3.0 software [14] which is based on the FALC algorithm developed by A. W. Harris *et al*. [15] and the *Asteroid Light Curve Analysis* software developed by P. Pravec *et al*. [3]. To select the comparison stars, we used the *Comp Star Selector* module in *MPO Canopus* with the CMC-15 catalog which provides *Sloan r* and 2MASS *JHK* magnitudes.

Some asteroids have been analyzed in post-opposition phase. Thus, it is not possible to calculate the absolute magnitude $H$ directly. Ten light curves have been obtained using the reduced magnitude $HR$, which is the magnitude calculated at the observed phase angle and a hypothetic geometry assuming that the asteroid is at 1 au from Sun and 1 au from Earth [14] (please note that the reduced magnitude could be considerably different from the observed apparent magnitude). Other two light curves have been obtained based on the *derived magnitude*, which is based on a hybrid method in which *MPO Canopus* uses the instrumental magnitude difference between the target and the comparison star, then adding the catalog magnitude of the comparison star. Finally, three light curves have been obtained using the classic differential photometry method.

To fit the light curves, we use a prudence principle, preferring an approximate result instead of a bad result. Typically, for fitting the curves, we used orders 3 to 5, depending on the data quality, choosing order 3 when the quality is low – for example when the *Star B Gone* step fails to work properly, or when the signal to noise (S/N) is low [16]. The quality of the linkage between sessions is important while selecting the fit order. In case the session linkage is good between many observed nights (resulting in lower dispersion data), then we used order 5 fits.

### 3.3. ASTEROID SHAPE CALCULATION

Based on the derived photometric data and the albedo (known from the literature or assumed based on spectral class), we have calculated some basic physical properties of the asteroids (size and shape). First, the size (equivalent diameter, assuming spherical shape) has been calculated using the following formula as a function of the absolute magnitude and albedo [17]:

$$D = \frac{1328}{\sqrt{\rho}} 10^{-0.2H},\qquad(1)$$

where $H$ is the absolute magnitude, defined as the magnitude of the hypothetical asteroid located at 1 au from Sun and 1 au from Earth and at zero phase angle (collinear Sun, Earth and asteroid), and $\rho$ is its albedo, defined as the ratio between the reflected and received Solar illumination. If the albedo is unknown, then we are assuming a standard value based on the asteroid spectral type.

Besides the equivalent diameter, another aim of this paper is to get the preliminary shape of the observed NEAs, based on the light curve analysis. For simplicity, we have assumed the basic shape asteroid model defined by three axis $a > b > c$. This model has been explained in details by Vincenzo Zappala [18, 19, 20] and makes the following three assumptions: the asteroids is a 3-axial ellipsoid body, it rotates around its shortest axis $c$ in its equatorial plane, and shows a well-defined light curve (showing clear maxima and minima). Using this method, we calculated the three-dimension shape model for all observed NEAs, presenting their shapes in the insets of Figure 1. In case of (433) Eros and (1980) Tezcatlipoca we include our derived model as well as the literature model resulted from radar observations.

## 4. RESULTS

In this section we present the observed targets, along with brief orbital data and any published photometric information (available rotation period) from the literature. To keep the reference list short, we ask the readers to search references lists in the online NEODyS-2 database [10]. In Fig. 1 we plot our derived light curves.

### 4.1. (433) EROS

This is the first discovered NEA, by G. Witt from Berlin in 1828, and second largest known NEA. It has become very well characterized *via* photometric, spectroscopic and radar observations, thus representing an opportunity to compare with our findings. Its class is Amor, moving in a close orbit ($a$ = 1.46 au, $e$ = 0.22, $i$ = 11°, MOID = 0.1476 au).

It is the largest of our sample (equivalent diameter 23.3 km) and it has a published period of 5.270 h [10], matching exactly our derived period using the small 0.36 m telescope.

### 4.2. (1866) SISYPHUS

This NEA was discovered by P. Wild at Zimmerwald in 1972, being classified as an Apollo ($a$ = 1.89 au, $e$ = 0.54, $i$ = 41°, MOID = 0.1021 au) [10]. It is considered a binary based on radar observations [21] that show a spike in the echo power spectrum, confirmed later by light curve photometry of Stephens *et al.* [22] who reported for the satellite an orbital period of 25.5 h. Assuming binarity, the primary has a period of about 2.4 h and an amplitude that ranges from 0.01 to 0.15 magnitudes. Its rotation period is P = 2.400 h, the absolute magnitude is $H$ = 12.510 mag and the diameter is 6.86 km [10].

Based on one night using our 0.36 m telescope, we reduced a period of P = 2.402 ± 0.002 h which agrees very well with previous work, and we got the reduced magnitude to $HR$ = 16.0 ± 0.03 mag. Its taxonomic class is S and the albedo $\rho$ = 0.372 [10] based on which we calculate an equivalent diameter of 7.215 km.

### 4.3. (1980) TZCATLIPOCA

This NEA was discovered in 1950 by A. G. Wilson and A. A. E. Wallenquist at Palomar, being classified of Amor type ($a$ = 1.71 au, $e$ = 0.36, $i$ = 27°, MOID = 0.2444 au) [10]. It was observed photometrically by many authors, among which B. Skiff *et al.* [23] who resolved its rotation period to $P$ = 7.2523 h, absolute magnitude H = 13.960 mag and diameter of 6.00 km [10].

Based on 4 nights using our 0.36 m telescope, we reduced a period $P$ = 7.23 ± 0.01 h, which is very close to published work. The taxonomic class is SU; Sl, S and the albedo $\rho$ = 0.128 [10] based on which we confirm the equivalent diameter of 6.00 km.

### 4.4. (4055) MAGELLAN

This NEA was discovered in 1985 by E. F. Helin, being classified as an Amor ($a$ = 1.82 au, $e$ = 0.3, $i$ = 23.2°, MOID = 0.2385 au). It was observed photometrically very recently by B. Warner *et al.* [24] who resolved its rotation period to $P$ = 7.496 h and absolute magnitude $H$ = 14.45 mag. Its taxonomic class is V, the albedo $\rho$ = 0.330 and equivalent diameter is 2.781 km [10].

Based on 4 nights using our 0.36 m telescope we reduced a period $P$ = 7.477 ± 0.003 h, very closed to previous results.

### 4.5. (7350) 1993 VA

This NEA was discovered in 1993 by R. H. McNaught at Siding Spring, being classified of Apollo type ($a$ = 1.36 au, $e$ = 0.4, $i$ = 7°, MOID = 0.0809 au). It was observed photometrically recently by B. Warner *et al.* [25] who derived a period of 3.580 h, absolute magnitude $H$ = 17.30 and equivalent diameter of 2.363 km.

Based on only two nights using our 0.36 m telescope, we reduced a period $P$ = 3.932 ± 0.024 h, and reduced magnitude $HR$ = 17.0 ± 0.07 mag. Its taxonomic class is C;X [2] albedo $\rho$ = 0.05 [10] based on which we calculate a diameter of 2.157 km.

### 4.6. (10150) 1994 PN

This NEA was discovered in 1994 by G. J. Garradd at Siding Spring, being classified of Amor type ($a$ = 2.38 au, $e$ = 0.5, $i$ = 46°, MOID = 0.2266 au). It was observed photometrically very recently by B. Warner who resolved its period P = 2.965 h [11, 12] and absolute magnitude $H$ = 15.3.

Based on 3 nights using the 0.36 m telescope, we reduced a period of $P$ = 2.966 ± 0.002 h, and reduced magnitude $HR$ = 16.6 ± 0.06 mag. Its taxonomic class and albedo remain unknown, thus its diameter could range between 2.6 and 5.9 km, assuming typical albedo of C and S type ($\rho$ = 0.04 and 0.20, respectively).

### 4.7. (35396) 1997XF11

This important PHA was discovered in 1997 by the Spacewatch survey, being classified of Apollo type ($a$ = 1.442 au, $e$ = 0.5, $i$ = 4°, MOID = 0.0004 au). It was observed photometrically in 2002 by P. Pravec *et al.* [10] who resolved its rotation period to $P$ = 3.2573 hand absolute magnitude $H$ = 16.770.

Based on 2 nights using the 0.36 m telescope, we reduced a period of P = 3.145 ± 0.002 h and reduced magnitude $HR$ = 15.5 ± 0.07 mag. Its taxonomic class is Xk; K [4, 5] but the albedo is unknown, so we calculate a diameter between 1.3 and 2.8 km.

### 4.8. (68216) 2001CV6

This NEA was discovered in 2001 by the LINEAR survey, being classified of Apollo type ($a$ = 1.32 au, $e$ = 0.3, $i$ = 18°, MOID = 0.0248 au). It was observed photometrically very recently by B. Warner *et al.* [10] who resolved its rotation period $P$ = 2.497 h, absolute magnitude $H$ = 16.50 and diameter of 1.400 km.

Based on only two nights using our 0.36 m telescope, we confirmed the period of $P = 2.497 \pm 0.003$ h and reduced magnitude $HR = 16.0 \pm 0.14$ mag. Its taxonomic class is Sq and the albedo $\rho = 0.29$ [10], based on which we calculate an equivalent diameter of 1.236 km.

### 4.9. (68346) 2001KZ66

This PHA was discovered in 2001 by the NEAT survey at Haleakala, being classified as an Apollo ($a = 1.51$ au, $e = 0.4$, $i = 17°$, MOID = 0.0396 au). It was observed with radar in 2006 and suggested to be a contact binary by Benner *et al.* [26] and very recently photometrically by B. Warner [25, 11] who solved its rotation period to $P = 5.63$ h.

Based on data mining of the Pan-STARRS archive, P. Veres derived an absolute magnitude $H = 17.5$ mag. Its taxonomic class is S but the albedo is unknown, so the estimated diameter is 1.3–2.3 km. Based on only 3 nights using our 0.36 m telescope, we confirmed the period $P = 5.633 \pm 0.002$ h.

### 4.10. (85989) 1999JD6

This PHA was discovered in 1999 by the LONEOS survey at Lowell Observatory, being classified of Aten type ($a = 0.88$ au, $e = 0.6$, $i = 17°$, MOID = 0.0481 au). During few oppositions, it was observed photometrically by a few authors including B. Warner *et al.* who resolved its rotation period to $P = 7.667$ h, the absolute magnitude $H = 17.1$ mag and diameter of 1.84 km [10].

Based on 3 nights using the 0.36 m telescope, we reduced a period of $P = 7.65 \pm 0.01$ h which is very close to previous work. Its taxonomic class is K; L; Cg, the albedo $\rho = 0.075$ [10], so we calculate a diameter of 1.844 km, matching well the other authors.

### 4.11. (137805) 1999YK5

This NEA was discovered in 1999 by the LINEAR survey, being classified of Aten type ($a = 0.83$ au, $e = 0.6$, $i = 17°$, MOID = 0.1148 au). It was observed photometrically in 2008 by B. Skiff [23] and very recently by B. Warner [25] who resolved its rotation period to $P = 3.930$ h, absolute magnitude $H = 16.60$ and equivalent diameter 3.879 km.

Based on 6 nights using the 0.36 m telescope, we reduced the period of $P = 3.468 \pm 0.004$ h and reduced magnitude $HR = 18.560 \pm 0.01$ mag. Its taxonomic class is X; RQ and albedo $\rho = 0.027$ [10], thus we calculate a diameter of 3.871 km. Because one of our data sessions (2 Feb 2016) showed a drop in magnitude during 30% of the rotation period (covered for 23 hours and well visible in the solid black triangles to the bottom right in the plot in Fig. 1), we suggest binarity hypothesis. This should be confirmed during next apparitions.

### 4.12. (141354) 2002AJ29

This NEA was discovered in 2002 by the LONEOS survey, being classified of Amor type ($a = 1.99$ au, $e = 0.45$, $i = 11°$, MOID = 0.0892 au). Despite one very recent attempt of B. Warner [11, 12], its rotation period remains unknown, while its physical data is also unknown: diameter between 1.0 and 2.3 km assuming spectral class between C and S type (albedo $\rho = 0.04$–$0.20$), and absolute magnitude $H = 17.3$.

Based on 10 nights using the 0.36 m telescope, we reduced a period $P = 19.89 \pm 0.004$ h, and reduced magnitude $HR = 15.7 \pm 0.02$ mag. Based on the albedo limits, we calculate an affective diameter from 1.03 to 2.30 km.

### 4.13. (154244) 2002 KL6

This NEA was discovered in 2002 by the NEAT survey, being classified as an Amor ($a = 2.31$ au, $e = 0.5$, $i = 3°$, MOID = 0.0637 au). It was observed photometrically in 2010 by A. Galad [27] and very recently by B. Warner [11, 12] who resolved its absolute magnitude $H = 17.50$ and rotation period $P = 4.6063$ h.

Based on 4 nights using the 0.36 m telescope, we reduced a period of $P = 4.61 \pm 0.01$h and reduced magnitude $HR = 14.8 \pm 0.04$. Its taxonomic class is Q, Sq [10], but the albedo remains unknown, thus the diameter is between 0.95 and 2.13 km.

### 4.14. (163899) 2003 SD220

This PHA was discovered in 2003 by the LONEOS survey, being classified of Aten type ($a = 0.83$ au, $e = 0.2$, $i = 8°$, MOID = 0.0194 au). It is an extremely slow rotator, observed photometrically very recently by B. Warner [11], then A. Carbognani and L. Buzzi [29] who suggest its rotation period to $P = 285$ h and absolute magnitude $H = 15.50$ from 18 nights.

Based on 4 nights using our 0.36 m telescope, we reduced a period of $P = 173.4$ h, quite shorter than the previous value, thus future observations using longer time-span and coordinated campaign are recommended. Its taxonomic class is S/Sr but the albedo remains unknown, thus its diameter is between 1.0 and 2.3 km [10].

### 4.15. (331471) 1984 QY1

This NEA was discovered in 1984 by E. F. Helin and P. Rose at Palomar, being classified as a very elongated Apollo orbit ($a = 2.50$ au, $e = 0.9$, $i = 14°$,

MOID = 0.1340). It was observed photometrically very recently by B. Warner [11] who suggested a rotation period $P$ = 45.5 h. Following 36 nights working this NEA and 1,141 data points collected by the present first author, we reduced a period of $P$ = 45.90 ± 0.03 h (matching well Warner's result) and reduced magnitude $HR$ = 16.0 ± 0.04 mag. Its taxonomic class and also the albedo remain unknown, thus its diameter ranges between 2.3 and 5.1 km (assuming albedo for C and S type).

This object is a slow rotator extremely difficult to solve, mainly because it needs many weeks coverage, and also because its light curve does not appear to be typical bimodal. During the reduction process, we noticed that some observations show delay effects in the light curve and also different maximum at the same phase moment. Given this, we suspect (331471) to be a *non-principal axis rotator* (NPA) [30], whose light curve seems to be composed by two light curves. The Fourier series method used in the photometric analysis is a powerful technique for the period calculation, but it is limited in order to analyze NPA asteroids, being based on a single period. If tumbling nature was confirmed, this asteroid could have a movement classified as *short-axis mode* (SAM) [31] where $c$ stands for the short-axis ($a, b > c$).

### 4.16. (337866) 2001WL15

This NEA was discovered in 2001 by the NEAT survey, being classified of Amor type ($a$ = 1.99 au, $e$ = 0.5, $i$ = 7°, MOID = 0.0635 au). It was observed very recently photometrically by M. Hicks *et al.* [32] and B. Warner [11] who derived a low amplitude and rotation period to $P$ = 8.65 h and absolute magnitude $H$ = 18.7.

Based on 4 nights using the 0.36 m telescope, we reduced a period of P = 6.334 ± 0.004 h (quite different from previous first author) and reduced magnitude $HR$ = 11.7 ± 0.04 mag. Its taxonomic class is Sk but the albedo remains unknown, thus its diameter is between 0.5 and 1.2 km.

### 4.17. (385186) 1994 AW1

This PHA was discovered in 1994 by the K. J. Lawrence and E. F. Helin at Palomar, being classified as an Amor ($a$ = 1.11 au, $e$ = 0.1, $i$ = 24°, MOID = 0.0196). In 1997, P. Pravec and G. Hahn suggested this as a possible binary asteroid [33]. It was observed photometrically very recently by B. Warner and J. Oey [11] who solved its rotation to $P$ = 2.52 + 22.3 h (primary + possibly secondary) and absolute magnitude $H$ = 17.4.

Based on 3 nights using the 0.36 m telescope, we reduced a period of $P$ = 2.49 ± 0.01 h, which fits well previous observations. Its taxonomic class is Sa and the albedo type is mh, based on which one could derive a diameter of the two possible bodies of about 0.8 and 0.3 R km.

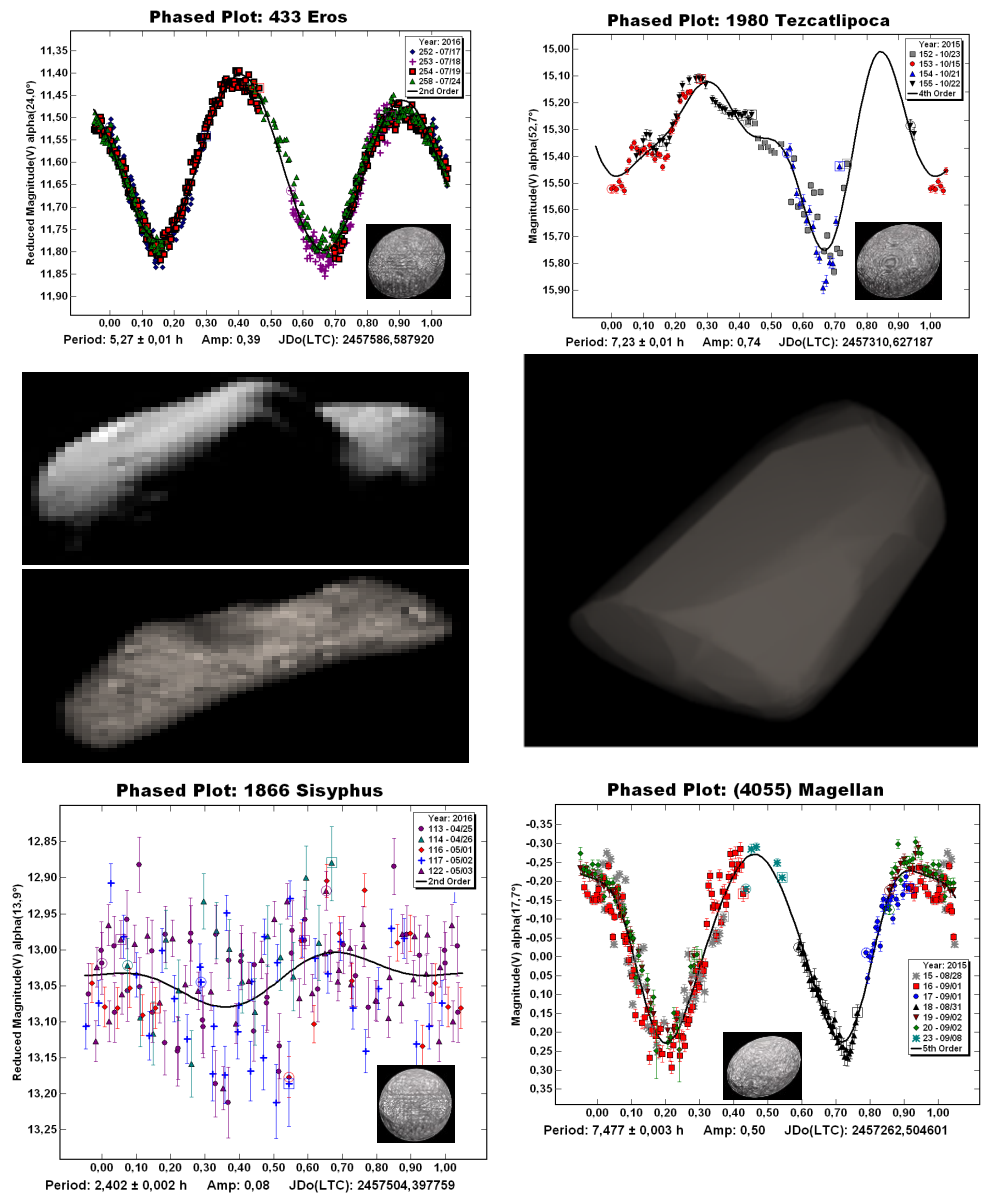

Fig. 1

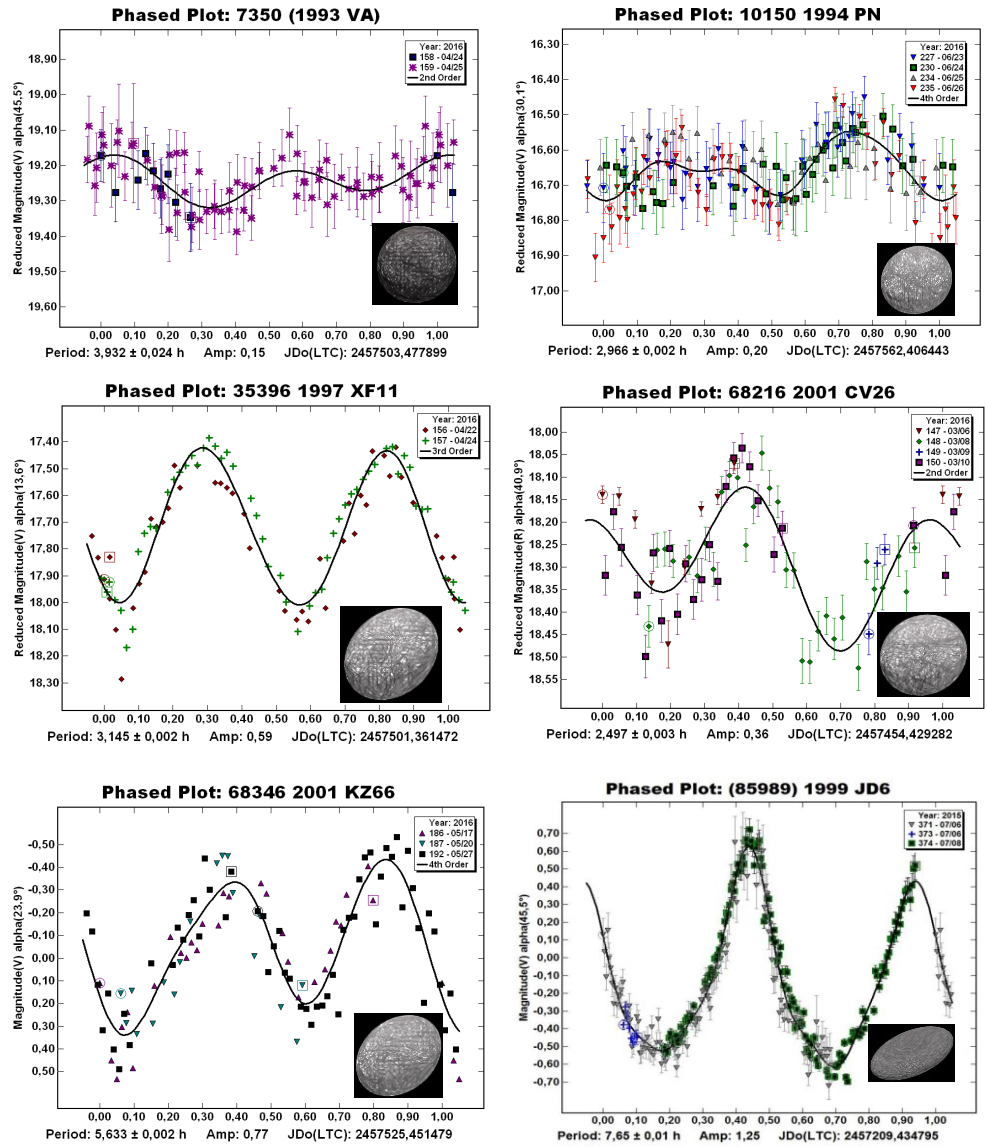

Fig. 1 (continued)

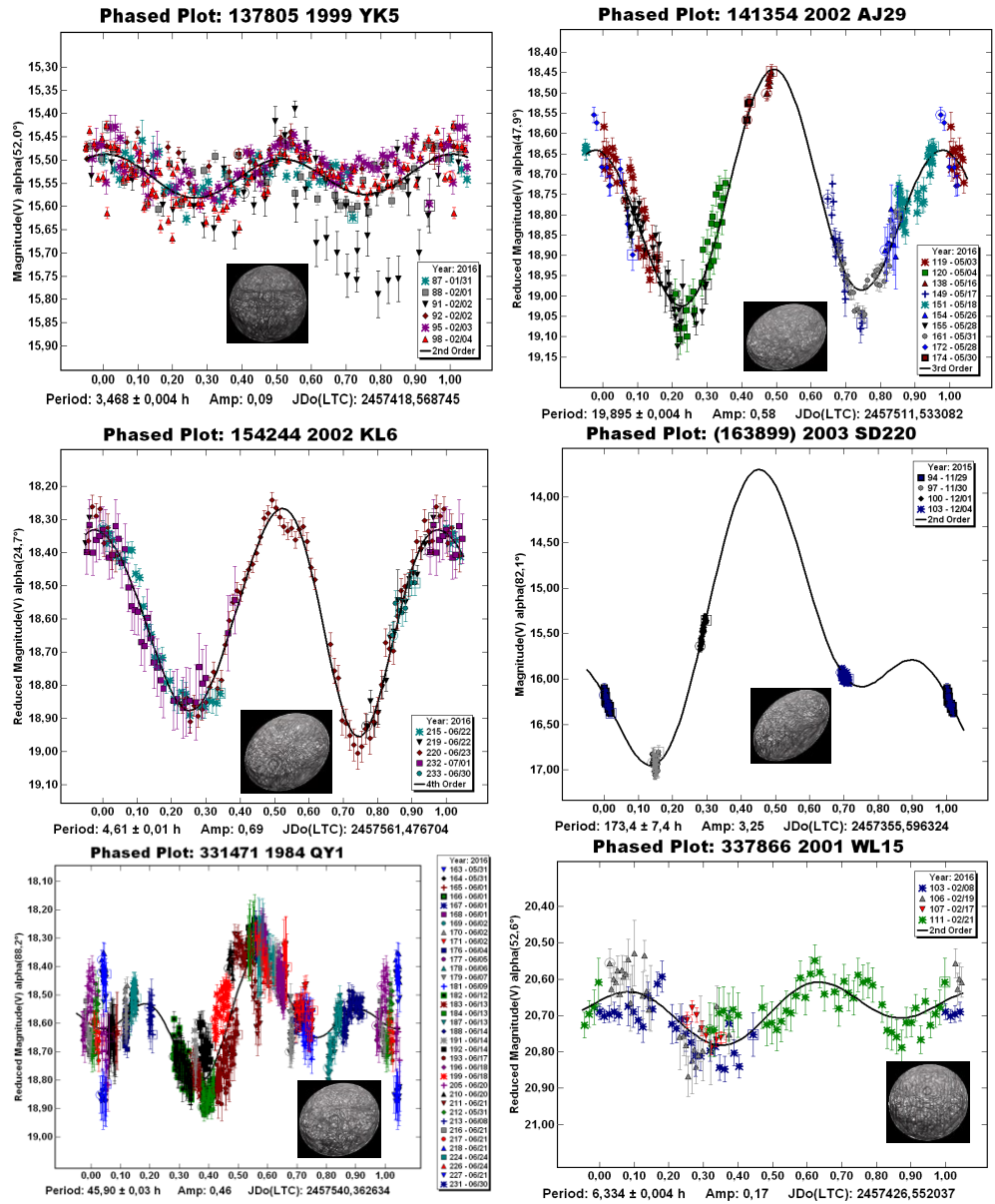

Fig. 1 (continued)

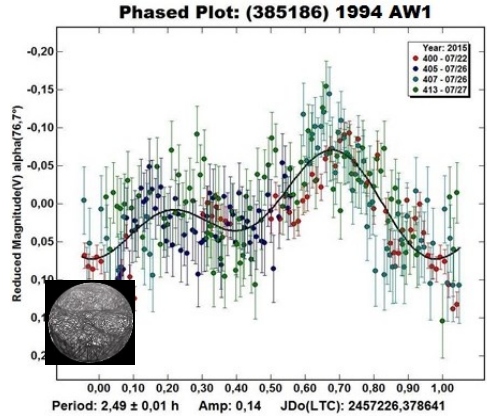

Fig. 1 (continued) – Light curves of the 17 NEAs observed and reduced in this paper.
In the inset we present our modeled 3-ellipsoidal shapes and comparison with literature models
in the case of (433) Eros and (1980) Tezcatlipoca.

## 5. CONCLUSIONS AND FUTURE WORK

During 2015 and 2016 we observed 17 light-curves of near Earth asteroids (NEAs) part of the EURONEAR efforts to study NEAs in Europe. This publication is the first part of a EURONEAR campaign which started in 2014 to survey NEA light-curves with other larger telescopes.

Most objects presented here were observed with the 0.36 m telescope owned by the first author (a new EURONEAR member) at his private Isaac Aznar Observatory in Spain, and one object in collaboration with another private telescope owned by the last co-author at Blue Mountain Observatory in Australia, thus we are proving that amateur astronomers could bring serious contributions in the physical characterization of NEAs.

Most rotations were independently solved based on observations carried at the same time (apparitions) by the American amateur Brian Warner and few other authors who sent their data to Minor Planet Centre and eventually published them in Minor Planet Bulletins. This way we could probe that 12 of our 17 targets match very well (all within 0.1 h in rotation periods) other results published independently. By comparison to other work, and using the same reduction software MPO Canopus (written by Brian Warner), and also by comparison with other data existent in the literature (including two NEAs targeted by radio), we could probe our data reduction approach and physical results, namely the rotation periods, reduced magnitudes and shape of the targeted NEAs.

One object has no period known previously, namely the asteroid (141354), which we observed during 10 nights, deriving $P = 19.895 \pm 0.004$ hours. Two other objects gave results quite different from published results (up to about twice in rotation period), namely (68346) and the very slowly rotator (137805), thus both will need re-observing in future apparitions. Another object (137805) is suspected to be binary and deserves refinement using larger telescopes. Finally, one other object, namely (331471) was very well covered by us (in 36 nights) and is suspected to be non-principal axis rotator and needs follow-up.

This work is part of the PhD thesis of the second author, co-advised by the third co-author of this paper. Part of this EURONEAR campaign to solve rotation periods of NEAs using smaller telescopes, in the future we plan to continue to use the 0.36 m Isaac Aznar telescope, other smaller telescopes accessible to the EURONEAR network in both hemispheres, and also to involve new students and the new small Observatory of the University of Craiova, as well as students from other Romanian universities and amateur astronomers for similar observations.